\documentclass[aps,prl,twocolumn,showpacs,10pt]{revtex4}%
\pdfoutput=1
\usepackage{amsfonts}
\usepackage{amsmath}
\usepackage{amssymb}
\usepackage{graphicx}%
\usepackage{rotating}
\usepackage{multirow}
\begin{document}
\title{Magnetic properties of single atoms of Fe and Co on Ir(111) and Pt(111)}
\author{C. Etz, J. Zabloudil and P. Weinberger}
\affiliation{Center for Computational Materials Science,  Getreidemarkt 9/134, A-1060
Vienna, Austria}
\author{E.~Y. Vedmedenko}
\affiliation{Institute for Applied Physics, University of Hamburg, Jungiusstr.~11, 20355
Hamburg, Germany}

\begin{abstract}
In using the fully relativistic versions of the Embedded Cluster and Screened
Korringa-Kohn-Rostoker methods for semi-infinite systems the magnetic
properties of single adatoms of Fe and Co on Ir(111) and Pt(111) are studied.
It is found that for Pt(111) Fe and Co adatoms \ are strongly perpendicularly
oriented, while on Ir(111) the orientation of the magnetization is only
out-of-plane for a Co adatom, for an Fe adatom it is in-plane. For comparison
also the so-called band energy parts of the anisotropy energy of a single
layer of Fe and Co
on these two substrates are shown. The obtained results are
compared to recent experimental studies using e.g. the spin-polarized STM
technique. \ 

\end{abstract}
\pacs{73.20.At, 72.10.Fk, 73.22.-f, 75.30.Hx, 73.20.Hb}
\maketitle

The potential application in non-volatile data storage devices is one of the driving forces
behind research into magnetic nanostructures. In state-of-the-art hard disk drives
a collection of a few hundred of single-domain particles (grains) are used to hold one bit
of information. If materials can be manufactured which exhibit sufficiently large anisotropies
and thermal stabilities it may become possible to store one bit in a single
grain~\cite{Gambardella}.
Such storage devices will require magnetic structures of precise atomic
arrangement, as -- if in addition the lateral dimensions of grains are further reduced -- the
influence of the perimeter atoms becomes increasingly important~\cite{Bode,Heinrich}
and, as is known, from previous studies the magnetic properties of each atom
in a nanostructure are highly influenced by its local environment~\cite{Gambardella,Bode,Heinrich,Corina}.

Using Scanning Tunneling Microscopy structures can be precisely tailored and
their magnetic properties determined.
In recent Scanning Tunneling Spectroscopy experiments~\cite{Hirjibehedin,Heinrich}
it has become possible to measure not only the Lande g-factor of individual atoms but also
their magnetic anisotropy.
The findings suggest that the anisotropy energy of a single atom may eventually be large enough
to use the magnetic state of an atom as a storage unit, pushing the ultimate limit for 
data storage density even further.
Since the magnetic properties of small clusters and single adatoms differ strongly from
those of bulk systems and even monolayers -- e.g. showing a much enhanced magnetic anisotropy
energy -- they do not only generate interest for their
technological relevance but also from a fundamental point of view.

In this paper we present a study of the magnetic moments and the angular
dependent band energy part of the magnetic anisotropy energy of single atoms
of Fe and Co, which -- in order to investigate the influence of different substrates --
have been deposited on Pt(111) and Ir(111). The calculations have been performed
by means of the Embedded Cluster
Method (ECM), a scheme based on the fully relativistic Screened Korringa-Kohn-Rostoker
(SKKR) method, in which we can treat impurities embedded into a
two-dimensional translationally invariant semi-infinite host. This approach
makes use of multiple scattering theory in which the electronic
structure of a cluster of embedded atoms is described by the so-called
scattering path operator given by the following Dyson
equation~\cite{Bence-1,kkr-book}:
%
\begin{equation}
\tau_{c}(\epsilon)=\tau_{h}(\epsilon)[1-(t_{h}^{-1}(\epsilon)-t_{c}%
^{-1}(\epsilon))\tau_{h}(\epsilon)]^{-1}\quad
\end{equation}
where $\tau_{c}(\epsilon)$ and $\tau_{h}(\epsilon)$ account for all the
scattering events within the embedded cluster and the host, $\mathbf{t}%
_{c}^{-1}$ and $\mathbf{t}_{h}^{-1}$ denote the single-site scattering
matrices for the "impurity" and for the host atoms, respectively. Once
$\tau_{c}(\epsilon)$ is known all corresponding local quantities, i.e., charge
and magnetization densities, spin and orbital moments, as well as the total
energy can be calculated.

In order to perform self-consistent calculations within local density
functional theory \cite{LDA}, for the calculation of the $t$-matrices and for
the multipole expansion of the charge densities (needed to evaluate the
Madelung potentials), a cutoff for the angular momentum expansion of $l_{\max
}=2$ was used. The potentials were treated within the atomic sphere
approximation (ASA). In all self consistent calculations the orientation of
the magnetization was chosen to point uniformly along the surface normal ($z$
axis). Structural relaxations of the cluster--substrate distance, which may
in principle affect the magnetic properties ~\cite{Pick,Lysenko,Shick,Guirado-Lopez}, have
been neglected. The host and the impurity sites refer to the positions of an
ideal \textit{fcc} lattice with the experimental lattice constants of Pt
($a=3.92$ \AA ) and Ir ($a=3.84$ \AA ). The self-consistent calculations were
performed using 102 $k_{\parallel}$-points in the irreducible Surface
Brillouin zone integrations and 16 energy points for the energy integrations
along a semicircular contour in the complex energy plane by means of a
Gaussian quadrature. To guarantee that all the perturbed host atoms with 
non-negligible influence on the calculated properties are taken properly into
account, we have increased the number of the
self-consistently calculated perturbed atoms around the adatom from 12 (the
first atomic shell around the impurity) up to 85 (the fourth atomic shell). It
should be noted that the perturbed atoms refer to substrate atoms and empty
spheres (part of the vacuum region).

In principle the magnetic anisotropy energy consists of two parts, namely the
difference in total energy $\Delta E_{\mu\nu}$ and in the magnetic
dipole-dipole interaction energy $\Delta E_{dd}$ (shape anisotropy) between
two given uniform orientations $\mu$ and $\nu$ of the magnetization. Since for
single adatoms the magnetic dipole-dipole interactions are of little
importance only $\Delta E_{\mu\nu}$ is considered, which in turn was
calculated by means of the force theorem as the corresponding difference in band
energies~\cite{Magnetic Force Theorem}, for details see Ref.~\cite{kkr-book}.
\begin{figure}[ptb]
\centering
\begin{tabular}
[c]{cc}%
\begin{tabular}
[c]{ll}%
{\includegraphics[scale=0.3]{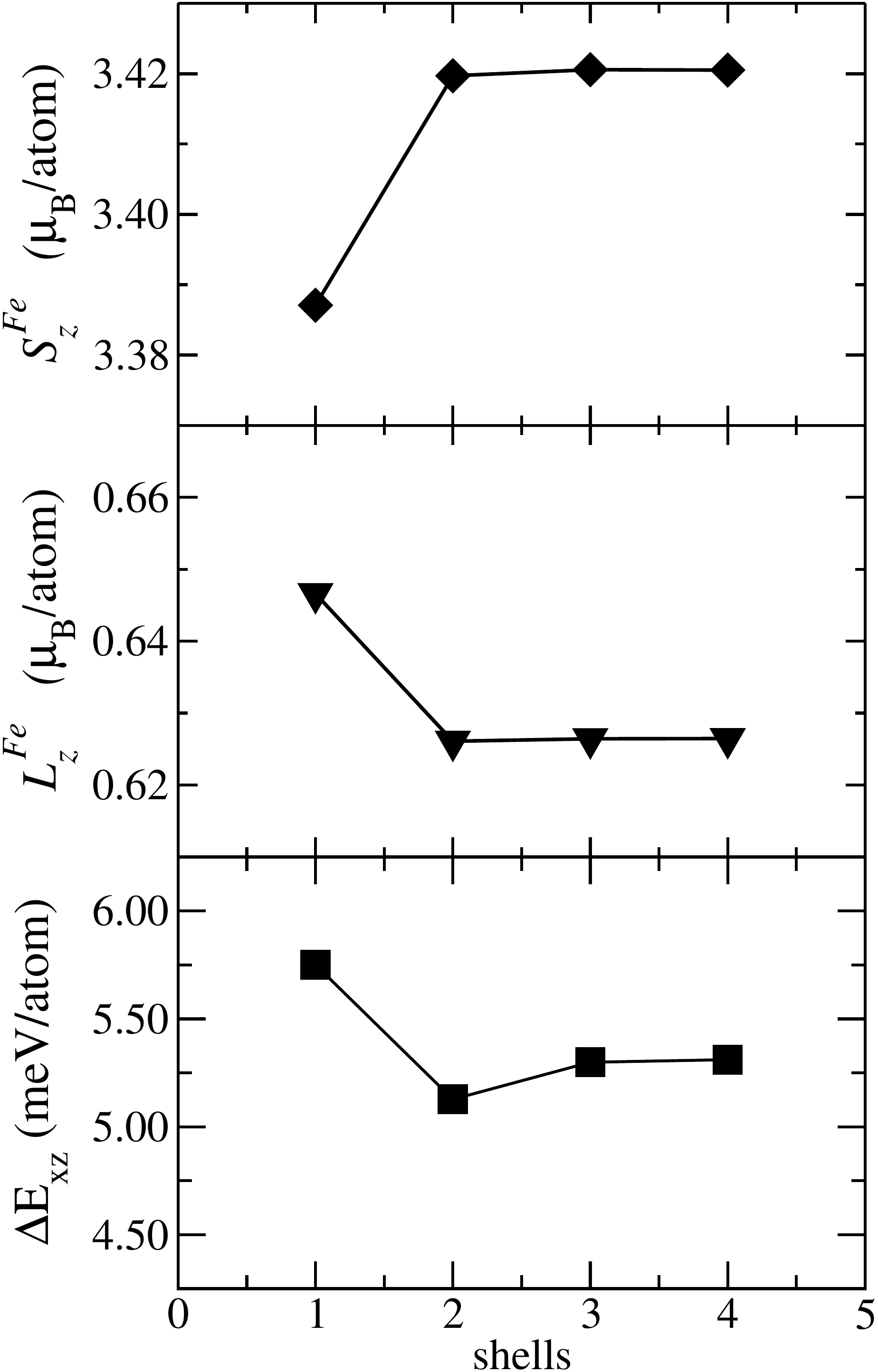}} &
\end{tabular}
&
\end{tabular}
\caption{Calculated spin (top) and orbital (center) magnetic moment, and
magnetic anisotropy energy (bottom) of a
single Fe adatom on Pt(111) as a
function of the number of self-consistently treated atomic shells around the
adatom.}%
\label{fig1}%
\end{figure}
\begin{figure}[ptbptb] \centering
\begin{tabular}[c]{cc}%
\begin{tabular}
[c]{ll}%
{\includegraphics[scale=0.18]{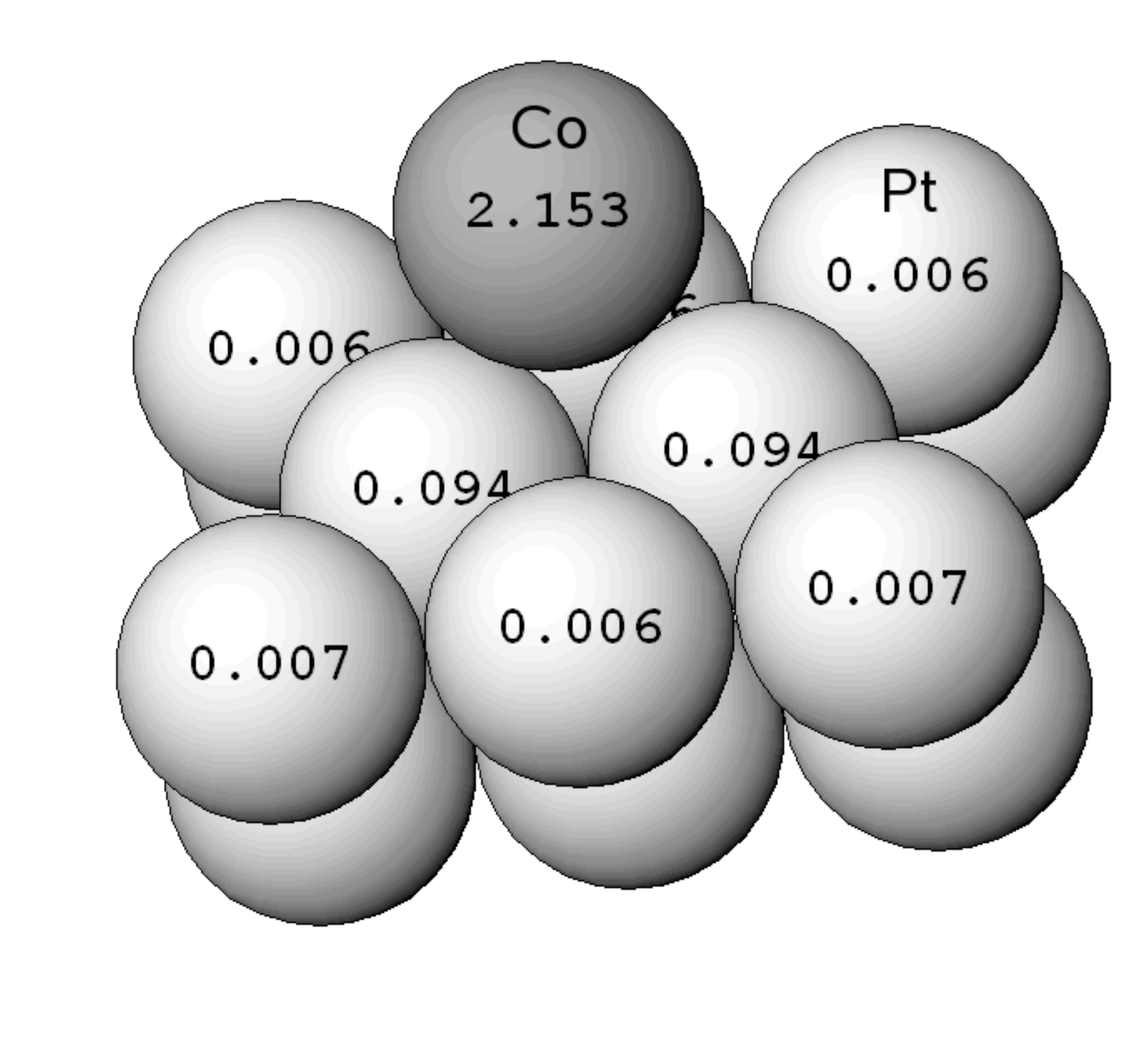}} &
{\includegraphics[scale=0.18]{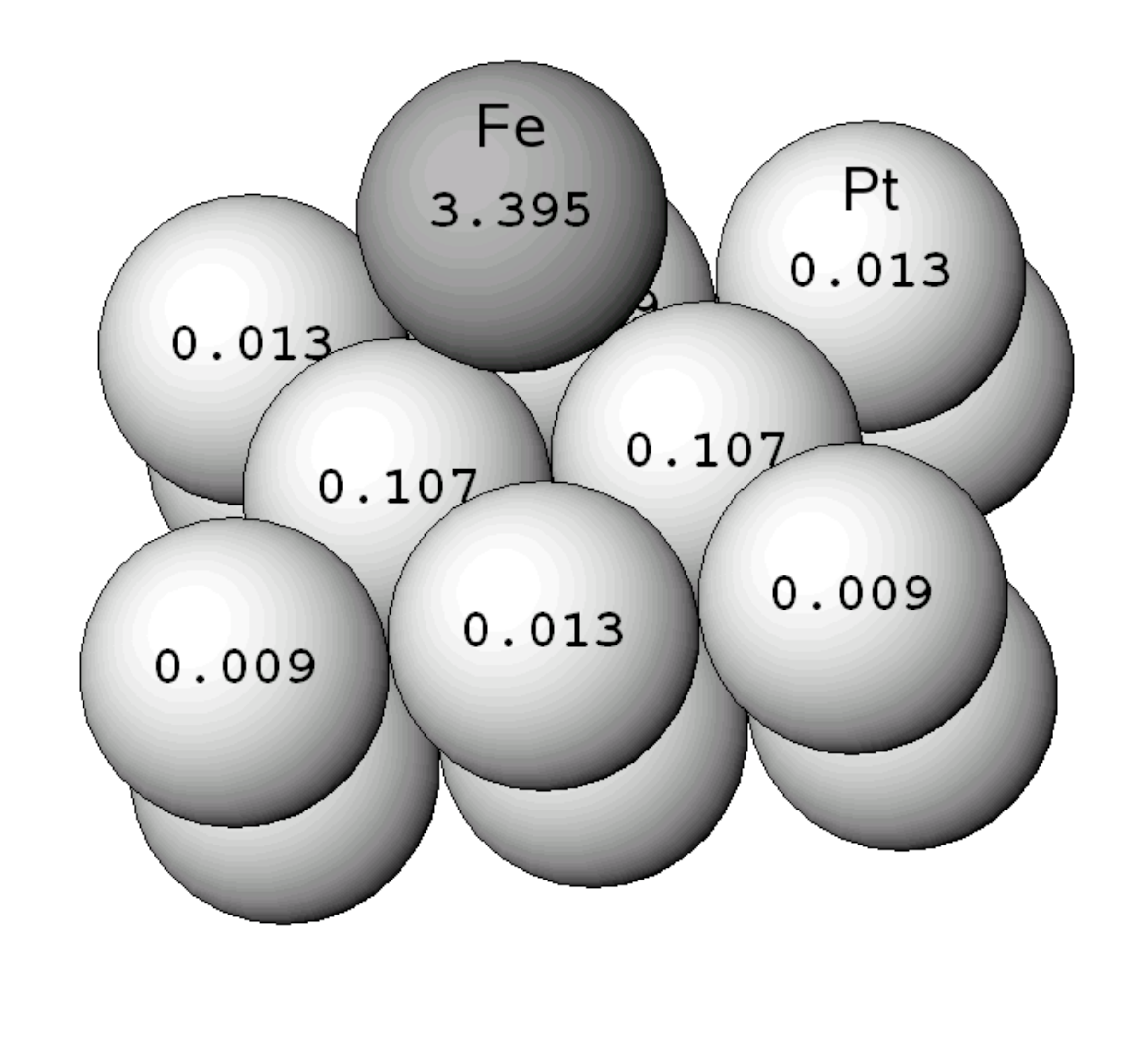}} \\
{\includegraphics[scale=0.18]{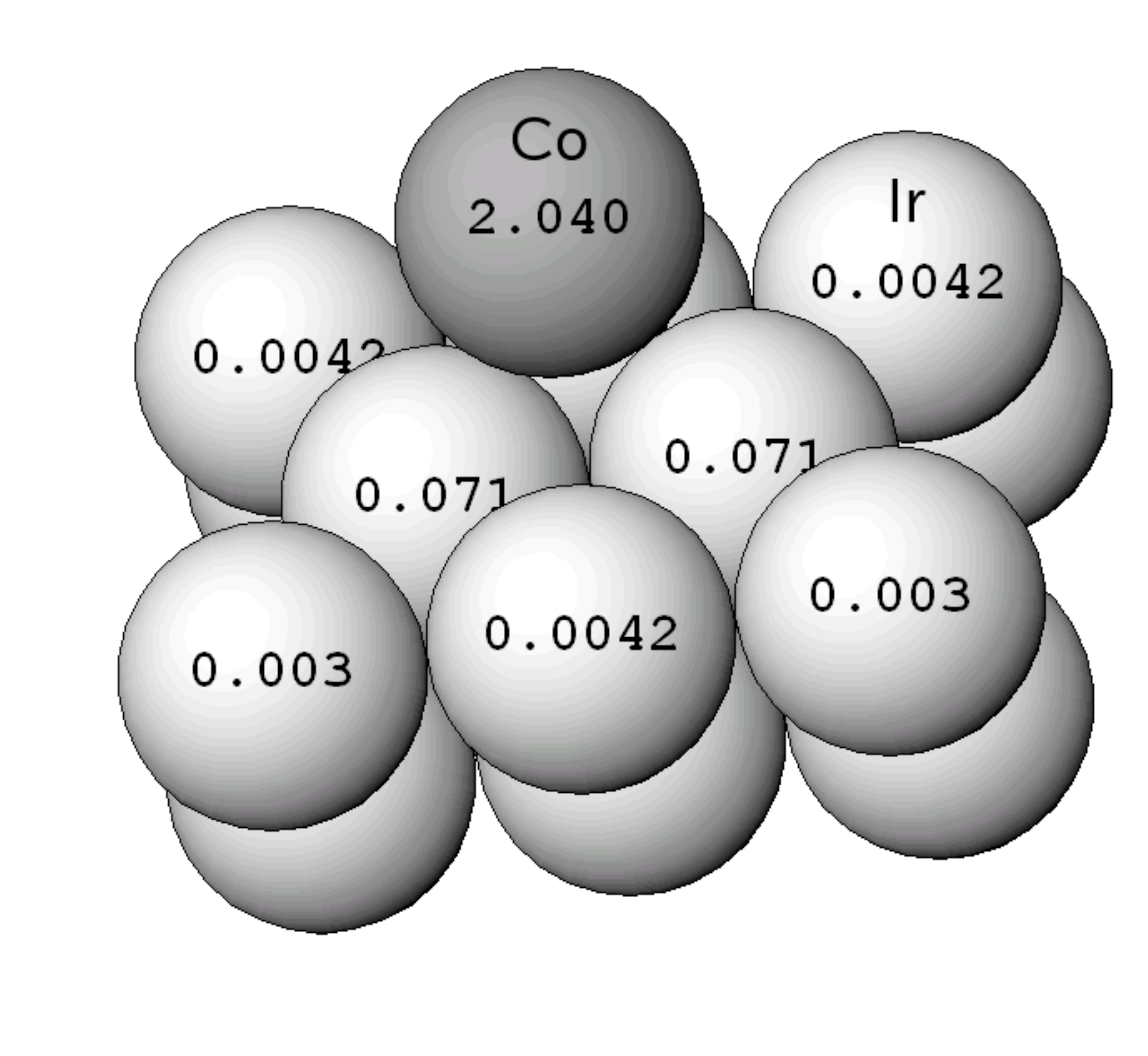}} &
{\includegraphics[scale=0.18]{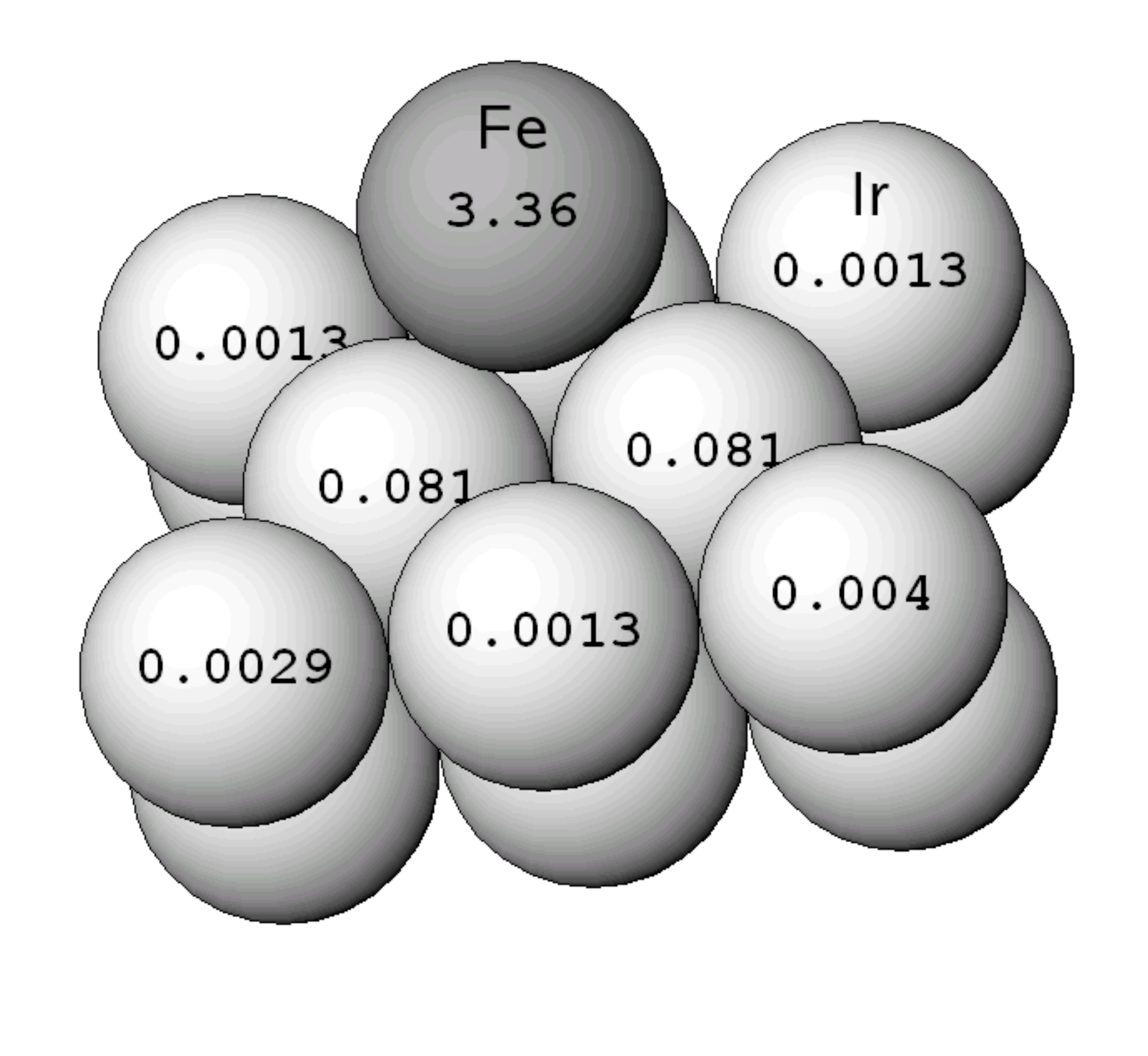}} \\
\end{tabular} &
\end{tabular}
\caption{Spin moments of the adatoms and induced spin moments in the topmost substrate layers.
The values correspond to the size of the moments along the easy magnetization axis, which is
out-of-plane in all cases except for the Fe/Ir system.}%
\label{fig2}%
\end{figure}
As can be seen from Fig.~\ref{fig1} reliable convergence of
the spin and orbital moments of the adatoms with respect to the number of
shells of neighbors used was obtained if only two shells of atoms were
taken into account. 
For the calculations of the magnetic anisotropy energy up to four shells of atoms were required to
obtain converged values as illustrated in Fig.~\ref{fig1}.

The size of the magnetic
moments of the adatoms exceed those of the respective bulk materials Fe and
Co, and of (complete) monolayers of Fe and Co deposited on Ir(111) and
Pt(111). This behavior, now quite well-known to be characteristic for small
magnetic clusters on top of metal substrates, \cite{Bence-1}, \cite{Corina} is due to the
lower coordination of the surface atoms which favors an incomplete quenching
of orbital contributions. For both kinds of adatoms, Fe and Co, the magnetic
moments are larger when deposited on a Pt substrate, a peculiar feature, which
most likely is caused by the stronger polarization of Pt than that of an Ir
substrate, see Fig.~\ref{fig2}. In fact the induced spin
magnetic moments in the nearest neighbour atoms of the Pt substrate are by about
0.02$\mu_{B}$ higher than for the Ir substrate, and the polarization rapidly decreases
by one order of magnitude for the second and third nearest neighbours.

The spin and orbital moments of the adatoms are summarized in Table~\ref{table1}.
At a first glance it can be realized that in relation to the bulk values the spin moments are
considerably increased. Compared to the (theoretical) 
value of bulk Co ($\mu_S^{\mathrm{hcp}}=1.6\mu_\mathrm{B}$)
the spin moments of the Co adatom
are increased by a factor of approximately 1.3 if deposited on either substrate.
For Fe, which has a bulk value of approximately $\mu_S^{\mathrm{bcc}}=2.1\mu_\mathrm{B}$,
this ratio is with 1.6 slightly larger.  While it is known that for 3d bulk systems, 
LSDA density
functional calculations predict the spin moments rather accurately (underestimating
the experimental values by only about 0.1$\mu_\mathrm{B}$), the orbital moments of Fe and Co, 
in particular, are off by about a factor of 2. Arguably, correlation effects may
play a prominent role in predicting the size of the orbital polarization correctly~\cite{imseok}.
The calculated values listed in Table~\ref{table1}
can therefore be expected to underestimate the actual size of the orbital moments of the adatoms.
However, it is worthwhile to consider the amount by which these values are increased compared to
the bulk values of LSDA calculations. The orbital moment of Co in bulk is $0.078\mu_\mathrm{B}$
and that of Fe $0.043\mu_\mathrm{B}$. Due to the reduced coordination and the different
chemical environment these values are increased for both an Fe and a Co adatom on the Ir substrate by
a factor of approximately 6.3.  In the case of a Pt substrate the Co and Fe orbital moments
are 9.3 and 14.6 times larger, respectively.
This increase of the orbital polarization is accompanied by an enhancement of the
magnetocrystalline anisotropy.

\begin{figure}[ptbptb] \centering
\begin{tabular}[c]{cc}%
\begin{tabular}
[c]{ll}%
{\includegraphics[scale=0.18]{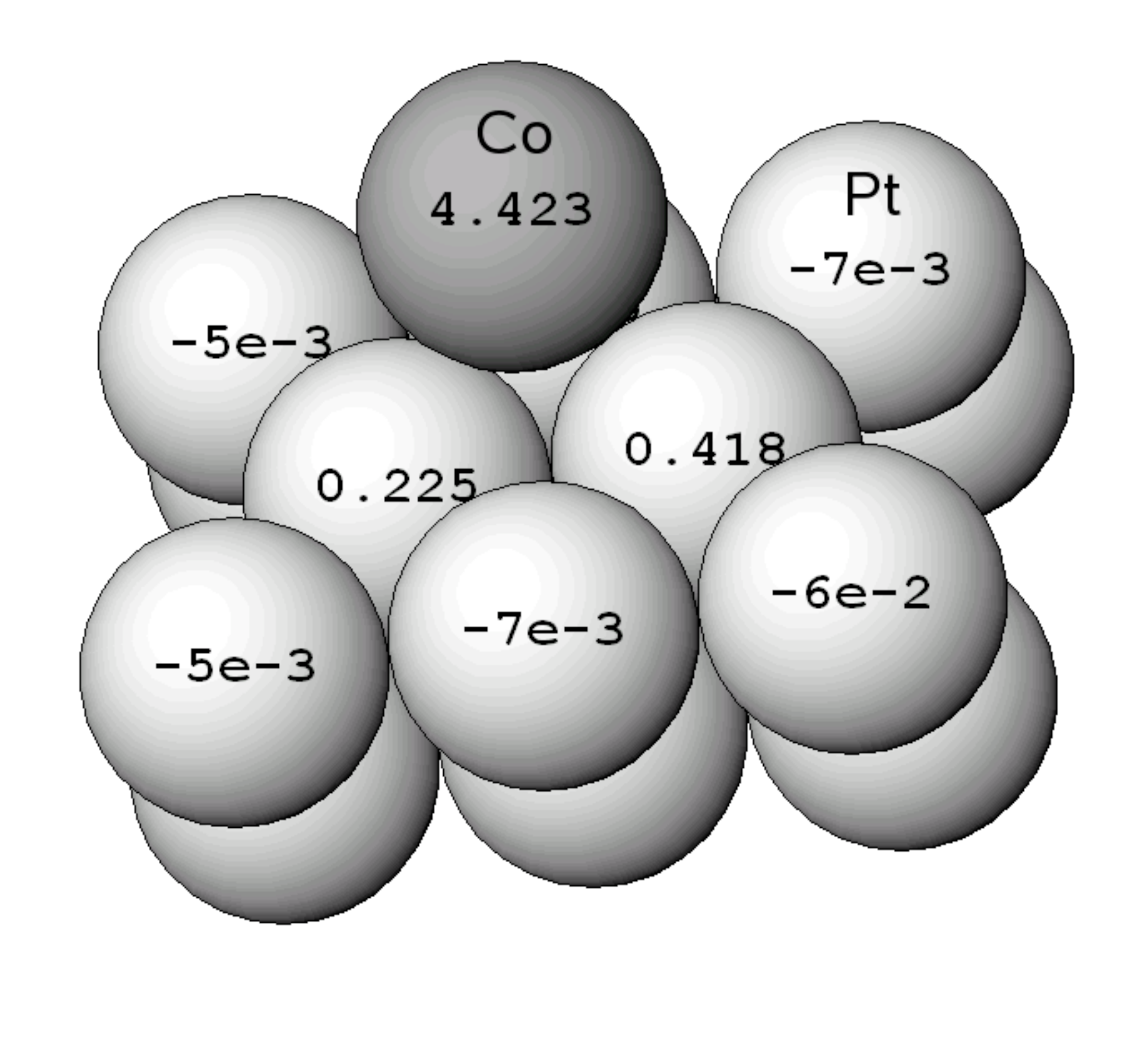}} &
{\includegraphics[scale=0.18]{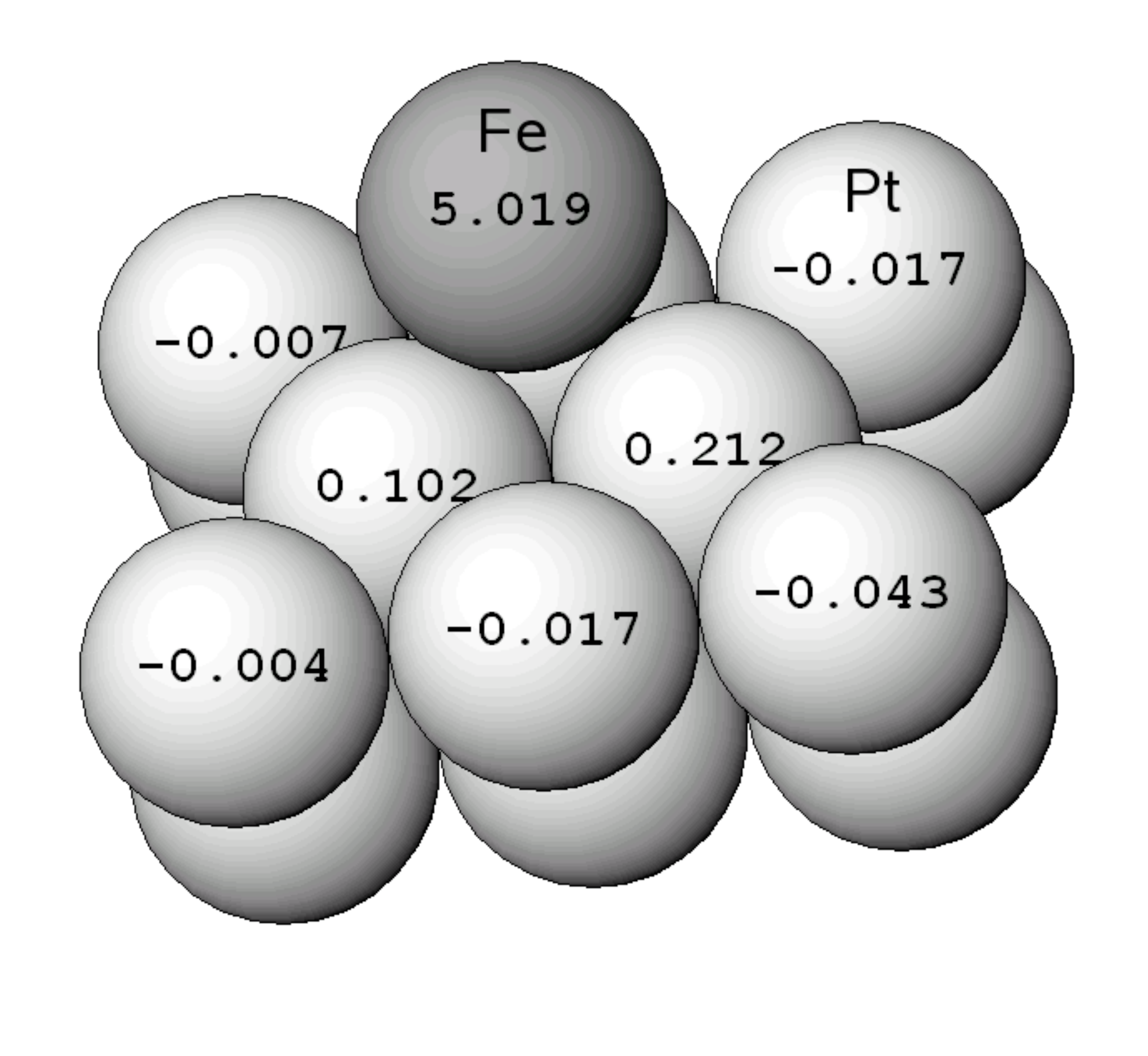}} \\
{\includegraphics[scale=0.18]{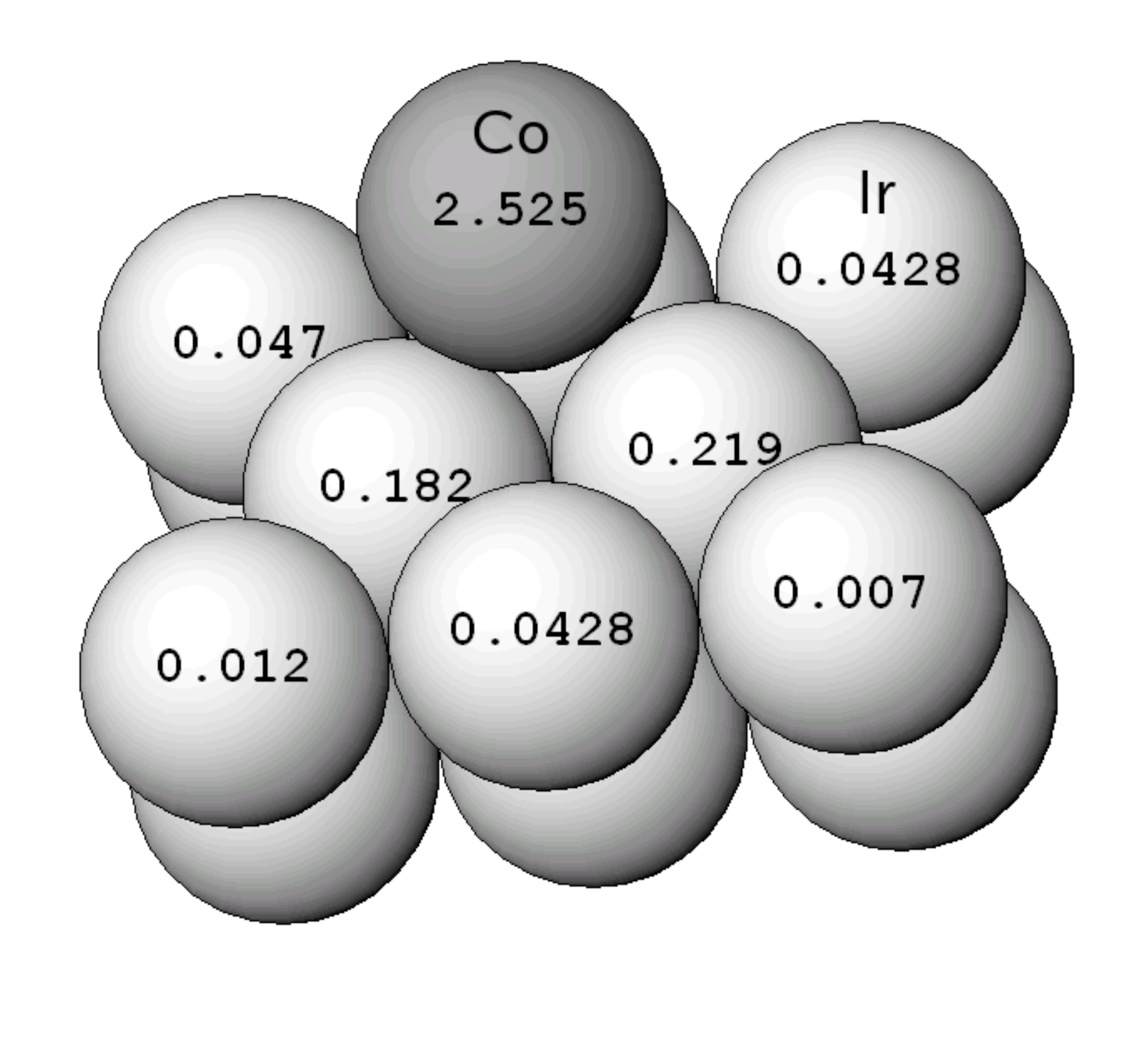}} &
{\includegraphics[scale=0.18]{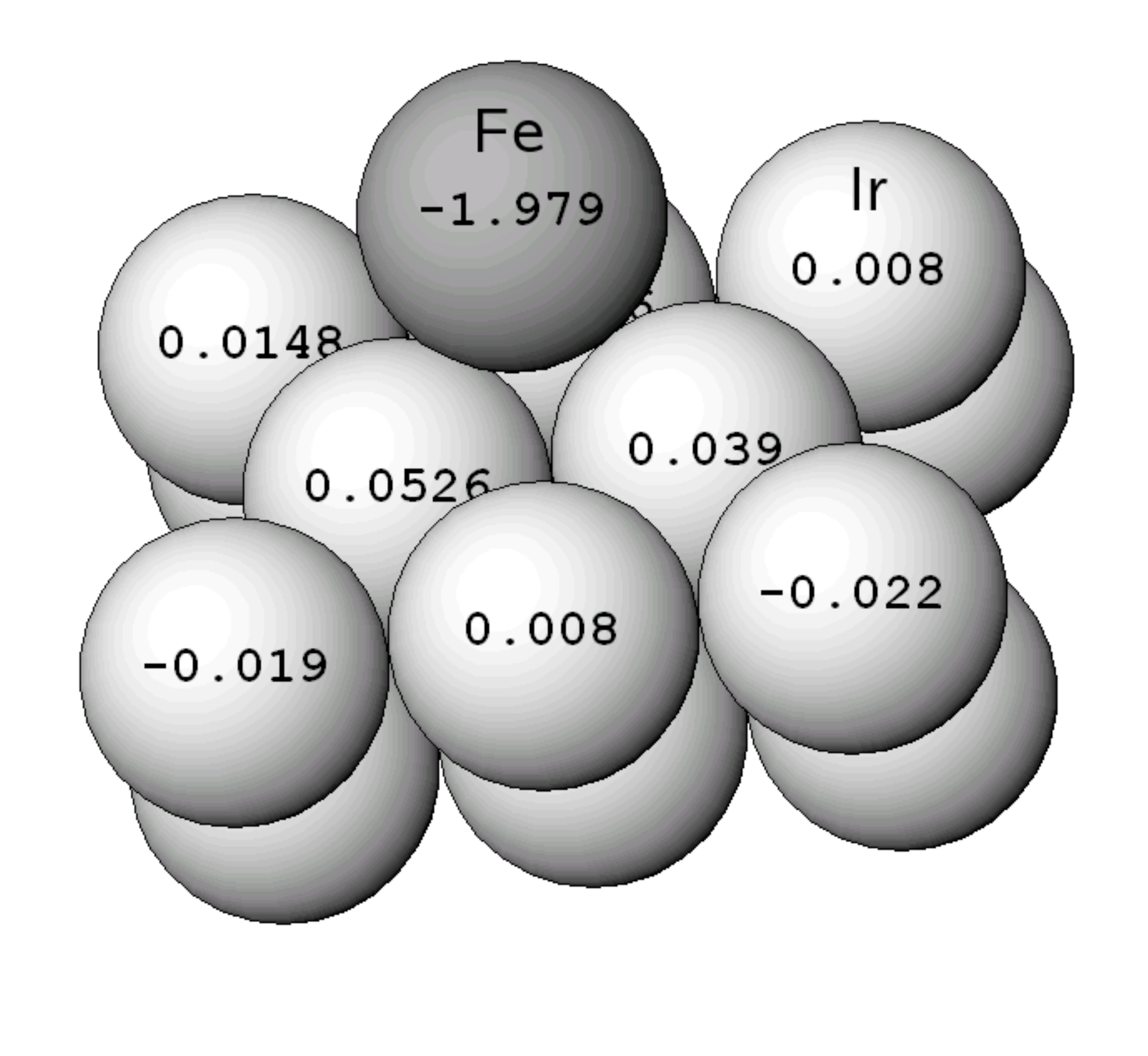}} \\
\end{tabular} &
\end{tabular}
\caption{Contributions to the MAE of the adatoms and the substrate. From left to right and
top to bottom: Co/Pt, Fe/Pt, Co/Ir, Fe/Ir}%
\label{fig3}%
\end{figure}

\begin{figure}[ptb]
\centering
\includegraphics[scale=0.4]{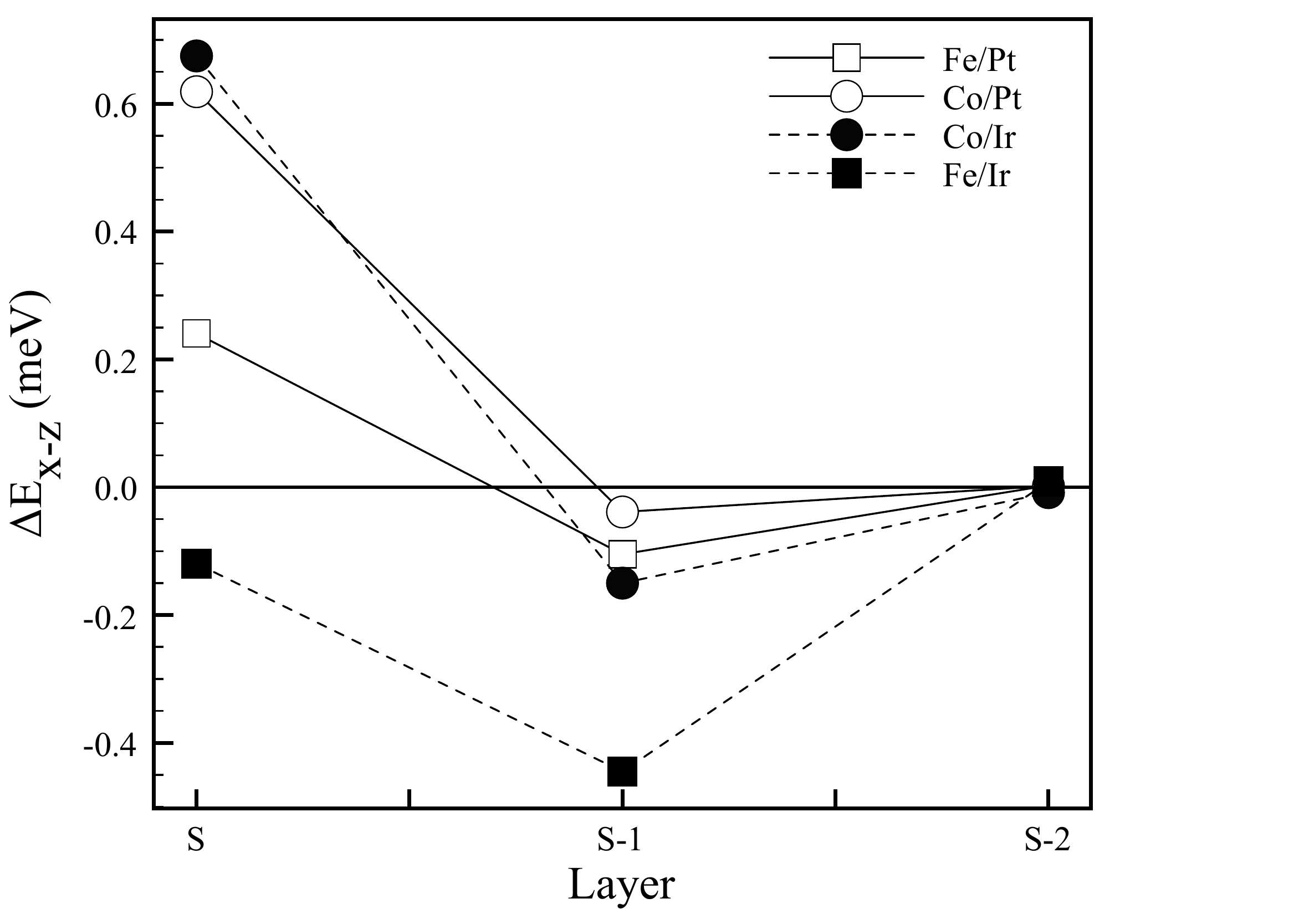}
\caption{Contributions of the substrate atoms in the surface layer (S)
and the two layers below (S-1, S-2) to the
total MAE.}
\label{fig4}
\end{figure}

\begin{figure}[ptb]
\centering
\begin{tabular}
[c]{cc}%
\begin{tabular}
[c]{ll}%
{\includegraphics[scale=0.3]{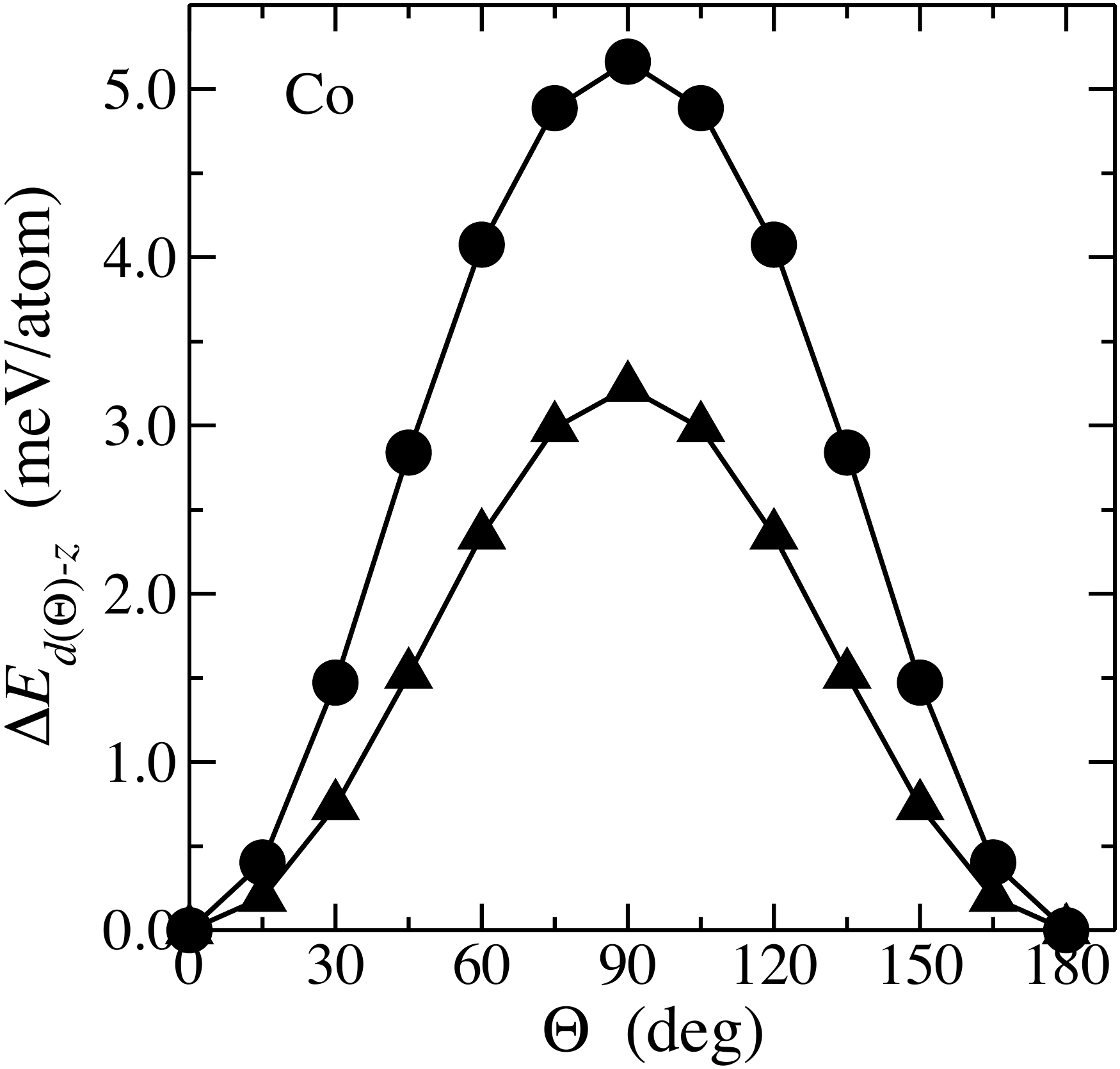}} &
\end{tabular}
&
\end{tabular}
\caption{Variation of the magnetic anisotropy energy of a single Co adatom on
Ir(111)(triangles) and on Pt(111)(circles) as a function of the orientation of the
magnetization with respect to \textit{z}-axis, as specified
by the polar angle $\Theta$.}
\label{fig5}
\end{figure}

\begin{figure}[ptb]
\centering
\begin{tabular}
[c]{cc}%
\begin{tabular}
[c]{ll}%
{\includegraphics[scale=0.3]{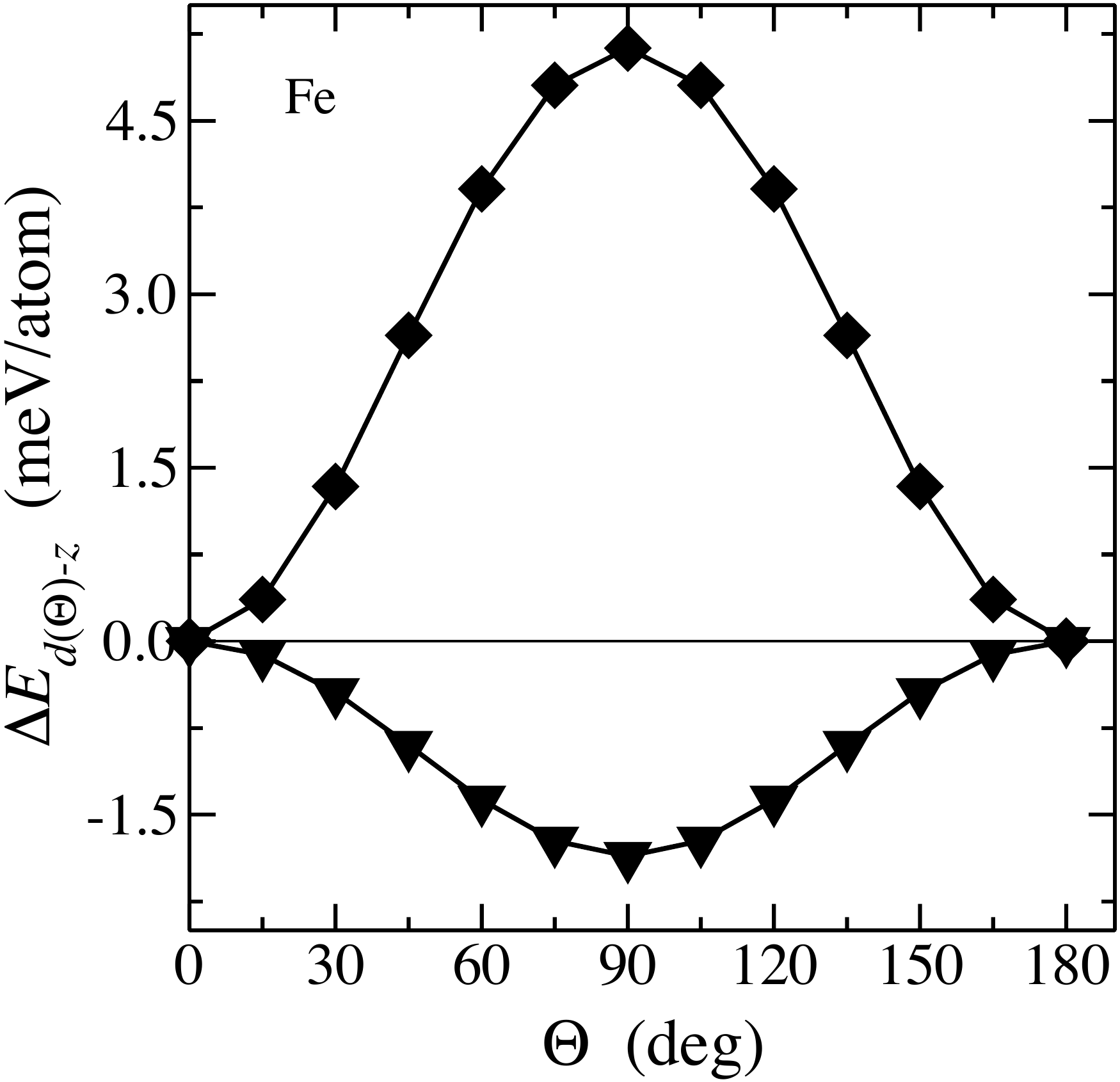}} &
\end{tabular}
&
\end{tabular}
\caption{Variation of the magnetic anisotropy energy of a single Fe adatom on
Ir(111)(triangles) and on Pt(111)(diamonds) as a function of the orientation of
the magnetization with respect to \textit{z}-axis, as specified by the polar
angle $\Theta$.}%
\label{fig6}%
\end{figure}


\begin{table}[ht]
\begin{center}
\begin{tabular}
[c]{| c | c | c || c | c | c | c | c |}%
\hline
\multicolumn{3}{|c|}{System} & $\Delta E_{xz}$ & $S_x$ & $S_z$ & $L_x$ & $L_z$ \\
\hline
\multirow{2}{*}{\rotatebox{90}{Pt(111)}} & \multirow{2}{*}{Fe} & ML & ~~-0.710~~ & ~~3.018~~ & ~~3.016~~ & ~~0.093~~ & ~~0.113~~ \\ \cline{3-8}
& & adatom & 5.310 & 3.514 & 3.395 & 0.266 & 0.628 \\ \cline{2-8}
&\multirow{2}{*}{Co} & ML & 0.123 & 1.987 & 1.988 & 0.117 & 0.147 \\ \cline{3-8}
& & adatom & 5.021 & 1.973 & 2.153 & 0.483 & 0.726 \\ \cline{2-8}
\hline
\multirow{2}{*}{\rotatebox{90}{Ir(111)}} & \multirow{2}{*}{Fe} & ML & -0.063 & 2.828 & 2.827 & 0.117 & 0.121 \\ \cline{3-8}
& & adatom & -2.655 & 3.359 & 3.341 & 0.267 & 0.243 \\ \cline{2-8}
& \multirow{2}{*}{Co} & ML & 1.395 & 1.893 & 1.900 & 0.126 & 0.142 \\ \cline{3-8}
& & adatom & 2.982 & 2.008 & 2.040 & 0.427 & 0.494 \\ 
\hline
\end{tabular}
\caption{Anisotropy energies [meV], spin and orbital magnetic moments $[\mu_B]$
of single monolayers of Fe and Co on Pt(111) and Ir(111) as compared to the corresponding
adatoms values.}
\label{table1}
\end{center}
\end{table}

Within our method it is possible to calculate the contributions of individual atoms
to the total MAE.  The values obtained are illustrated in Fig.~\ref{fig3}
for the adatom and atoms in the surface layer of the respective substrates, and
hence we can evaluate that portion of the anisotropy energy which is attributed to the substrate.
Interestingly we find a significant dependence of these contributions on the type of
deposited adatom. 
If Fe is deposited on Pt the substrate contributes only about 5.5\%, whereas in the case of
an Co atom the Pt atoms add 11.9\% to the total MAE. In contrast, the atoms of an Ir substrate
contribute 15.3\% for the Co adatom, and even 25.5\% if the adatom is Fe.
The latter case is the only instance when the preferred magnetization direction due to the MAE is
perpendicular to the surface normal. In that case, interestingly, the major contribution of the
substrate does not come from the atoms in the surface layer closest to the adatom, but from the
subsurface layer (S-1, c.f. Fig.~\ref{fig4}).

Independent of the substrate for a Co adatom $\Delta E_{\mu\nu}$ predicts
strongly an out-of-plane orientation of the magnetization with the easy axis
along $z$ direction, see Fig.~\ref{fig5}, while for an Fe adatom,
see Fig.~\ref{fig6}, the choice of the substrate seems to be significant: on top of
Pt(111) an out-of-plane orientation with an easy axis along $z$ applies, while
deposited on Ir(111) an easy axis along $x$ is preferred.

Finally for matters of comparison to the single adatoms $\Delta E_{\mu\nu}$ of
a complete Fe and a Co monolayer ferromagnetically
coupled \cite{afm} to the substrate was studied
within the framework of the fully relativistic SKKR
method \cite{kkr-book}. The results of this study are displayed in Table
\ref{table1}. 
One immediately observes that the orbital magnetic moments of the adatoms are
more sensitive to the chemical environment than the spin moments and in the meantime,
the orbital moment anisotropy is larger on the Pt substrate than on the Ir.
In comparing now Figs.~\ref{fig5} and \ref{fig6} with the values
in Table~\ref{table1}, one easily can see that $\Delta
E_{\mu\nu}$ is substantially larger for a Co adatom on Ir(111) or Pt(111) than
for the corresponding monolayer. In the case of Fe adatoms one even has a
reversed situation:\ according to our calculations and in a good agreement with experimental findings
\cite{Repetto} a single monolayer of Fe on Ir(111) or Pt(111)
exhibits an in-plane magnetic anisotropy
while a single Fe adatom on Pt(111) is strongly perpendicularly oriented.

{\bf An additional contribution to the magnetic anisotropy comes from the
magnetic dipole-dipole interaction energies (shape anisotropy), which is
always negative \cite{Lazi} and consequently favors an in-plane orientation of the
magnetic moments.
For single magnetic monolayers on metal substrates the shape anisotropy is
rather small (typically about -0.1 meV) and 
was not taken into account in the present calculations.
It should be noted that the
shape anisotropy becomes very important indeed whenever the number of magnetic
monolayers is increased and then very often is the cause for so-called
reorientation transitions, see for example the discussion in Ref.~\cite{Christoph}.
The numbers given in Table~\ref{table1} compare the band energy contribution to the
magnetic anisotropy energy of a monolayer with that of a single adatom.}

\begin{figure}[ptb]
\begin{center}
\includegraphics[scale=0.30]{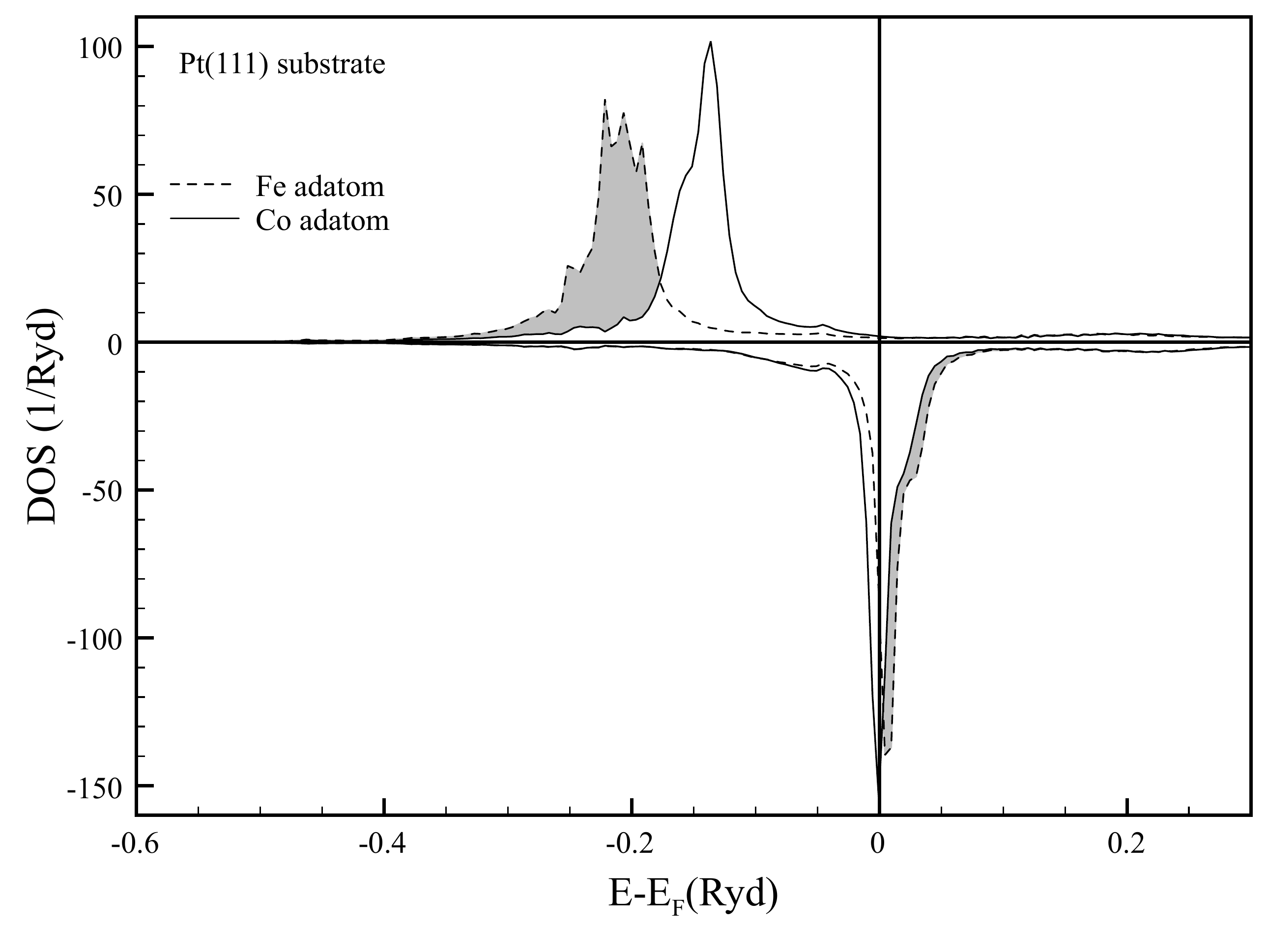}
\includegraphics[scale=0.30]{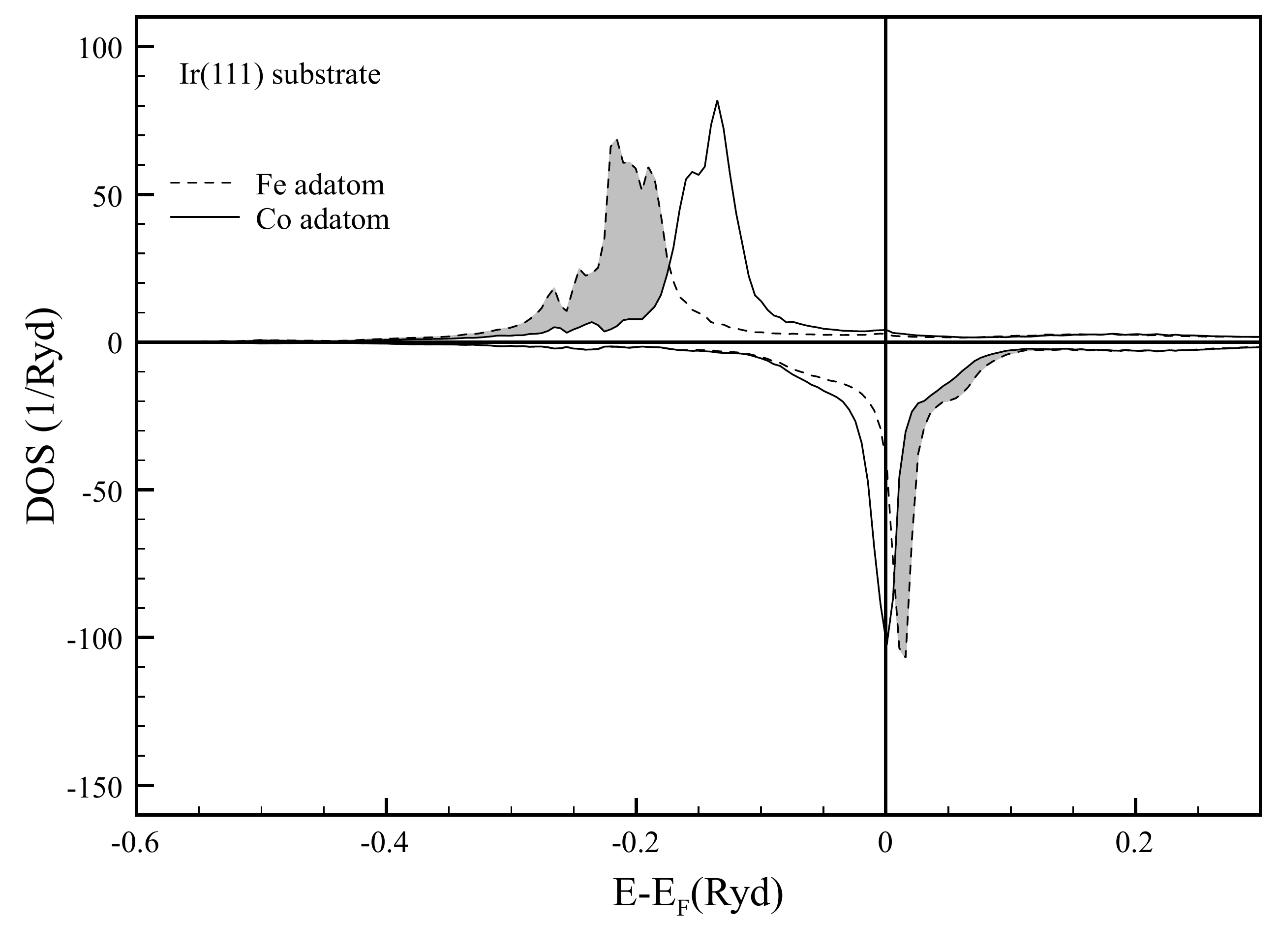}
\end{center}
\caption{``Spin-projected'' density of states of a single adatom of Fe (shaded area, dashed line)
and of a single Co adatom (black line) on Pt(111) (top) and Ir(111) (bottom). }%
\label{fig7}%
\end{figure}

{\bf Fig.~\ref{fig7} shows the total spin resolved density of states (DOS) for the Fe and Co adatoms
on the Pt and Ir substrate,
respectively. Note that the magnetic field was taken along the surface normal in all cases.
As can be expected the DOS appears to be very similar for both Fe and Co, with 
the spin-up DOS always completely filled. The spin-down DOS of Fe and Co are of almost identical
shape, but the peak is shifted slightly to lower energies for Co to accomodate the additional
electron.  Comparing the DOS on the Pt and Ir substrates, one can identify a broadening
of the DOS on Ir which results in the slightly lower spin moments, $S_z$, of Fe and Co
(c.f. Table~\ref{table1}).

As has been shown in Refs.~\cite{Bence-2} and~\cite{Soederlin} the increase in the orbital moments,
as compared to the bulk or the monoloayer cases, is caused by the difference in the filling of the 
$d_\alpha$ ($\alpha = xy,xz,z^2,yz,x^2-y^2$) orbitals.
}

\begin{figure}[ptbptb] \centering
\begin{tabular}[c]{cc}%
\begin{tabular}
[c]{ll}%
{\includegraphics[scale=0.25]{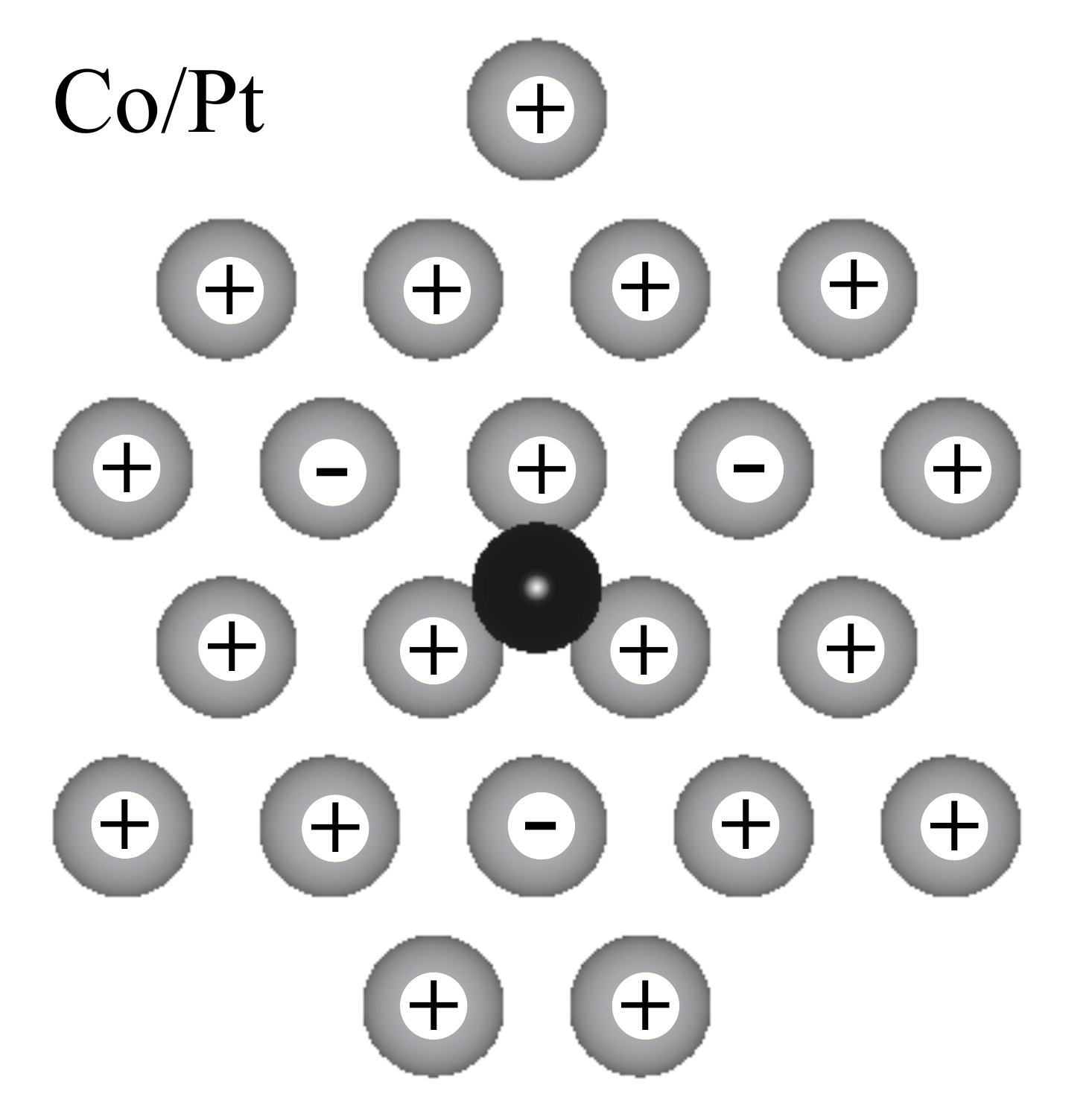}} &
{\includegraphics[scale=0.25]{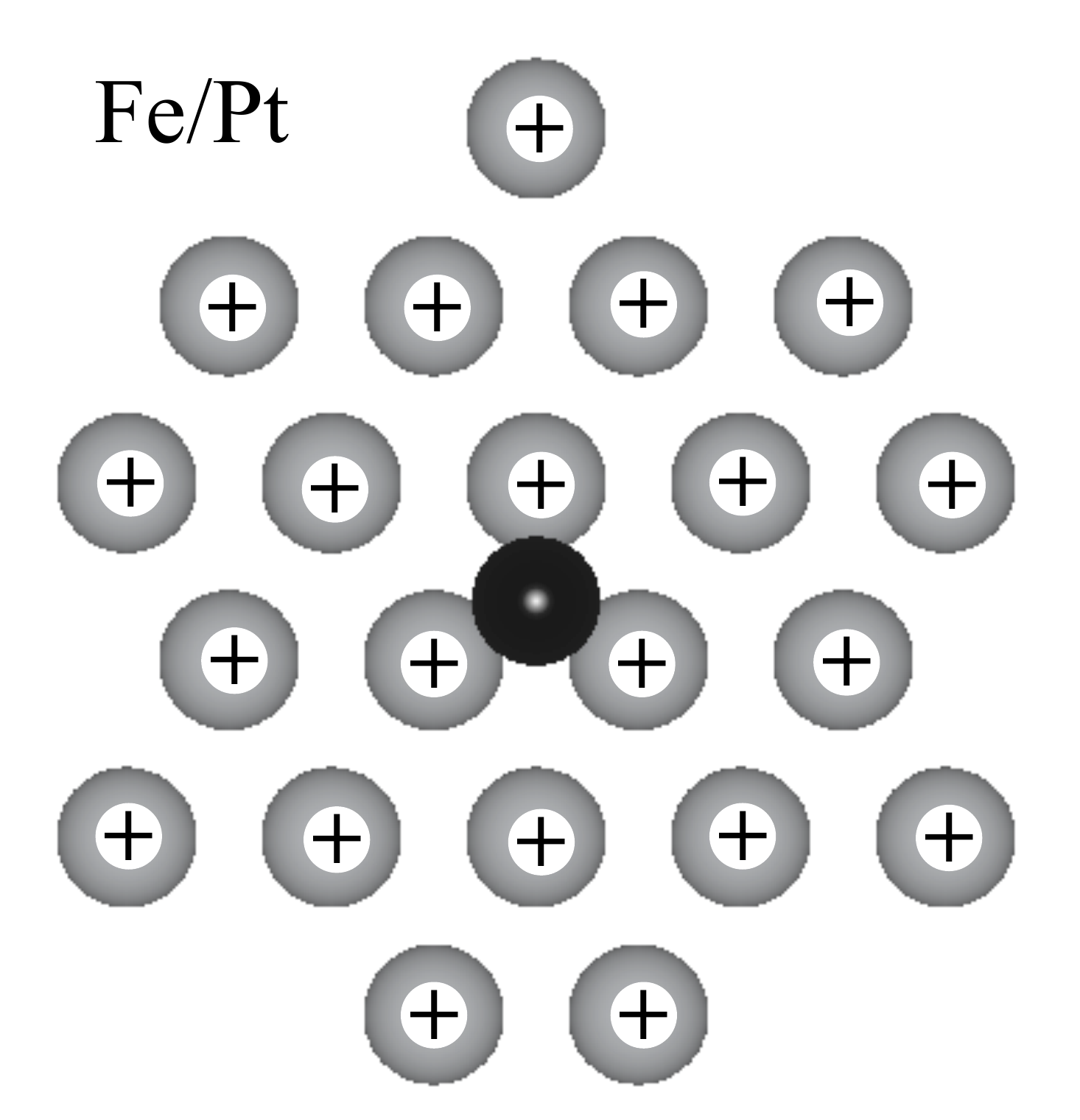}} \\
{\includegraphics[scale=0.25]{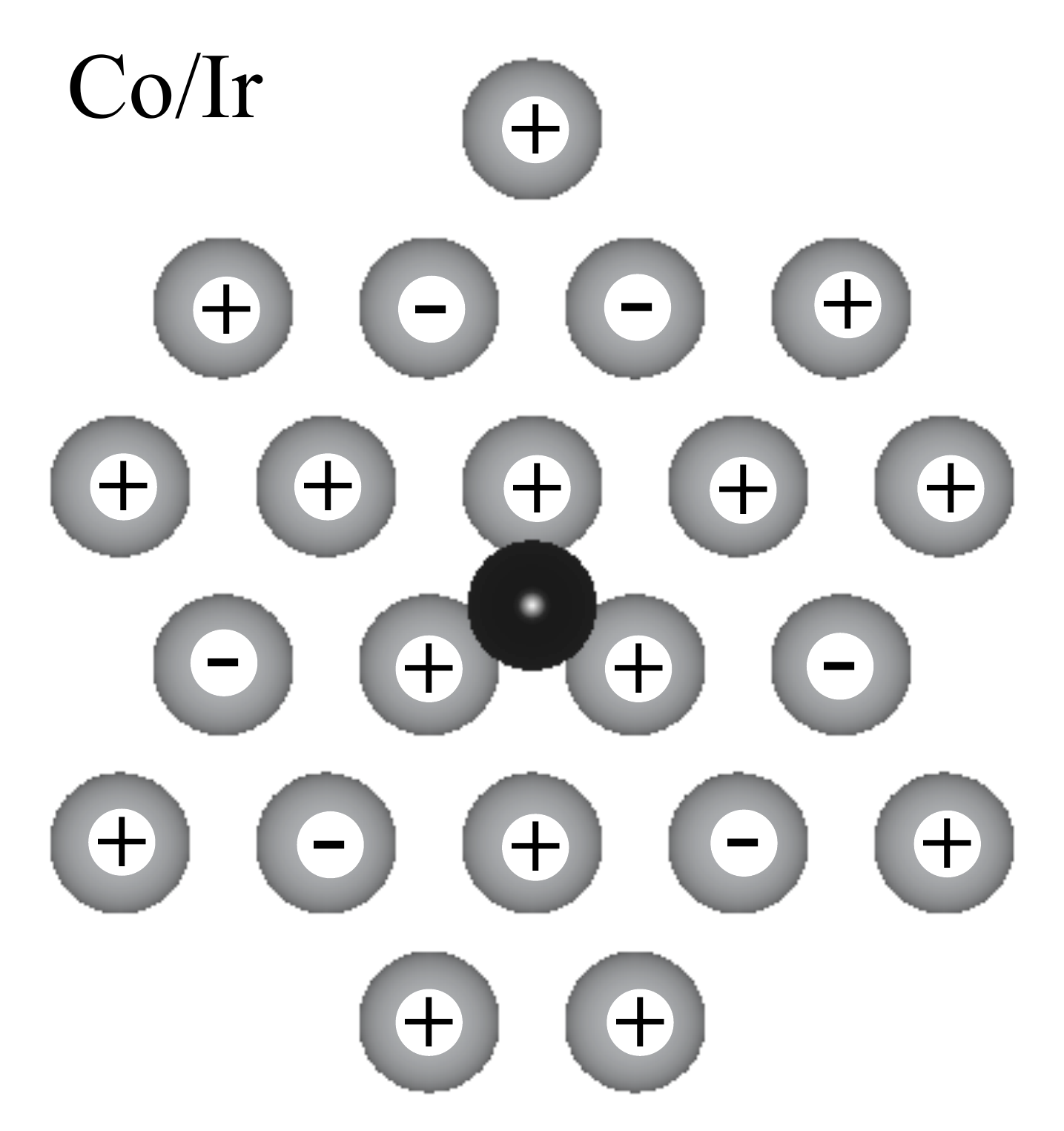}} &
{\includegraphics[scale=0.25]{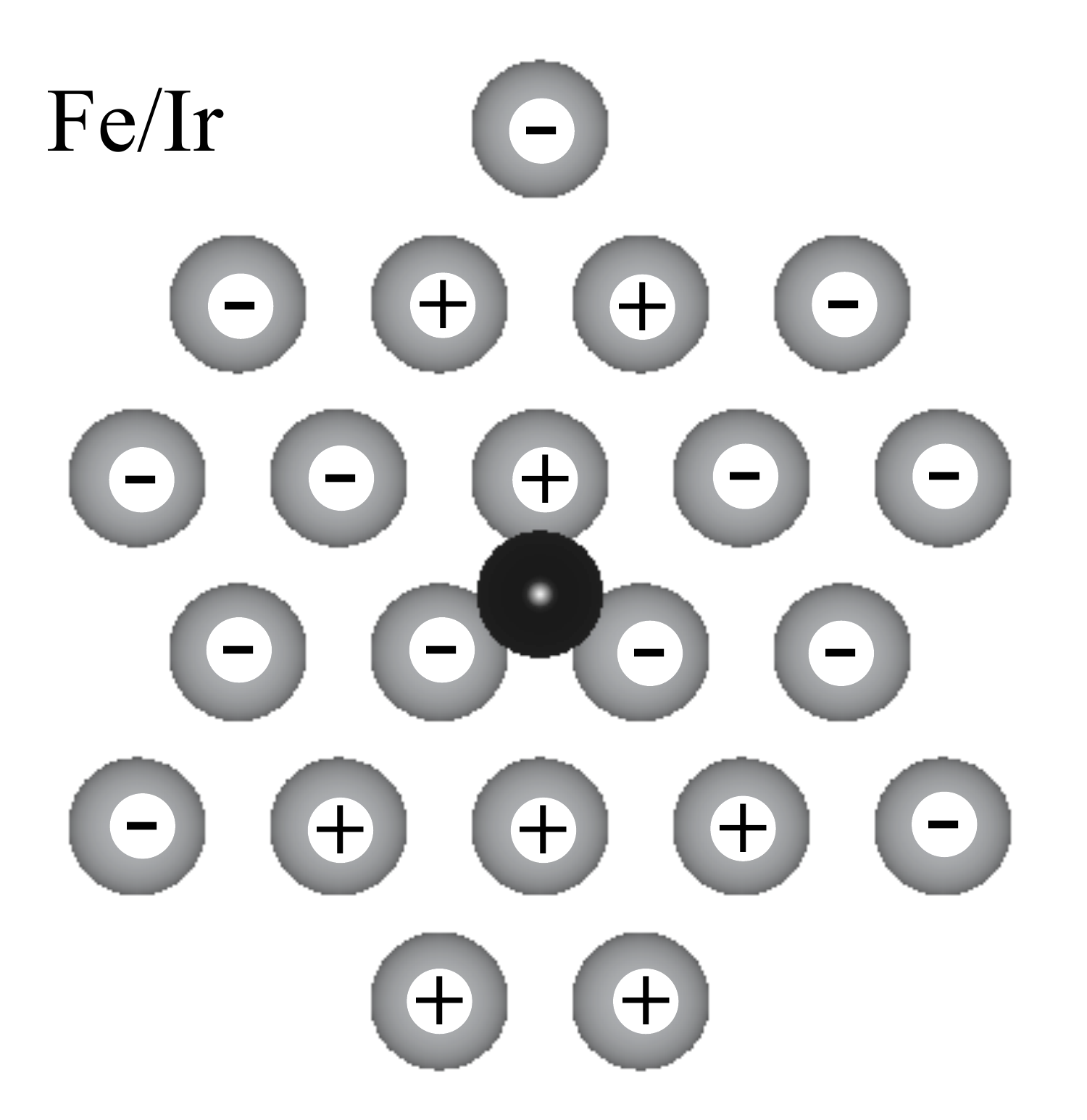}} \\
\end{tabular} &
\end{tabular}
\caption{Alignment of the spin and orbital moments in the surface layer
of the Pt and Ir substrate. The sign '+' means that
the induced moments are aligned parallel and '-' that they are aligned antiparallel.
The central, black, atom is the magnetic adatom.
From left to right and top to bottom: Co/Pt, Fe/Pt, Co/Ir, Fe/Ir}
\label{fig8}
\end{figure}

It has been noticed quite some time ago~\cite{wilhelm,tyer} that in interfaces of Fe films
with layers of the 5d elements the induced orbital moments can violate Hund's third rule.
As a consequence of this rule the spin and orbital moments should
align antiparallel for a less than half filled shell and parallel for a shell
more than half filled.  Even though strictly valid only for atoms
it seems that Hund's rules are also applicable to solids, however, with exceptions.
Since both Pt and Ir possess a more than half filled d-shell, $\mathbf{J=L+S}$ has to be expected. 
The relative orientation of spin and orbital moments are explored in Fig.~\ref{fig8} where
the atoms in the surface layer that occupy sites in the vicinity of the adatom are shown.
A '+' indicates parallel and a '-' antiparallel alignment of the moments. Note that the spin
moments align parallel with the spin moments of the adatom, as also illustrated in
Fig.~\ref{fig2}. However, this is not strictly the
case. Albeit not shown in this work, there are induced spin-moments in the subsurface layer
which align antiparallel to the spin of the magnetic adatom.
Fig.~\ref{fig8} shows, that in a Pt substrate the spin and orbital moment are parallel with Fe
on top, and only a few moments are aligned opposite with the Co adatom on top.
The situation is very much different for the Ir substrate, where many more atoms show antiparallel
alignments. The arrangement of the atoms exhibiting this anomaly shows a symmetry according
to the hexagonal 2D lattice in the case of the Co adatom. However, for the Fe adatom
this arrangement does not have the same symmetric pattern, because the easy magnetization direction
is in-plane along the x direction.
{\bf Small changes in the band filling have been shown to lead to such a behaviour~\cite{tyer}.}

Co nanostructures on Pt(111) were already studied in the past in terms of thin
films of Co$_{n}$ or (CoPt)$_{n}$ superstructures \cite{prb60-414-1999}, $n$
denoting the number of atomic layers or repetitions and in the form of finite
chains of Co atoms \cite{prb67-024415-2003,prb70-10040-2004}. In Ref.
\cite{prb60-414-1999} it was claimed that for a single layer of Co on Pt(111)
an in-plane orientation of the magnetization is preferred with a very small
anisotropy energy. In the spin dynamics study of Ref. \cite{prb67-024415-2003}%
, which is based on the same computational approach as used in here not only
the value of the magnetic anisotropy energy agreed very well with experiment
\cite{Gambardella}, but also the direction of the canted magnetization. A recent
study \cite{Meier} of the structure of a thin film Co/Pt$_{13}$/Co revealed
that an fcc-type stacking of Co was more favorable than an hcp-type stacking,
the interlayer distance between the Co and the first Pt layer being reduced by
10.1\% as compared to that of bulk Pt. Although a film with two magnetic
surfaces cannot be compared directly with a semi-infinite substrate coated
with a monolayer of a magnetic metal (in the case of a semi-infinite system
the Fermi energy is always that of the substrate, i.e., differs from that of a
thin film), these results indicate that layer relaxation might in special
cases be  important for investigating magnetic anisotropies.

Experimentally a study of Co nanostructures on Pt(111) \cite{Meier,Rusponi}
seems to lead to a rather complicated situation. In Ref.~\cite{Meier} an
out-of-plane magnetization of Co wires and islands is found and -- in order to
explain the measured domain wall width in the wires -- an effective anisotropy
constant between 0.08 and 0.17meV/atom for atoms within an island is proposed.
In Ref.~\cite{Rusponi} the edge atoms of small islands are made responsible
for their uniaxial out-of-plane magnetization. Of course none of these
experimental results can be compared directly with results for a smooth Co
monolayer on Pt(111) exhibiting two-dimensional translational symmetry. Finite
Co nanostructures (single adatoms or finite wires) on Ir(111) and on Pt(111), however, do show a strong
perpendicular anisotropy and support the experimental findings.

\begin{figure}[ptb]
\begin{center}
\includegraphics[scale=0.35]{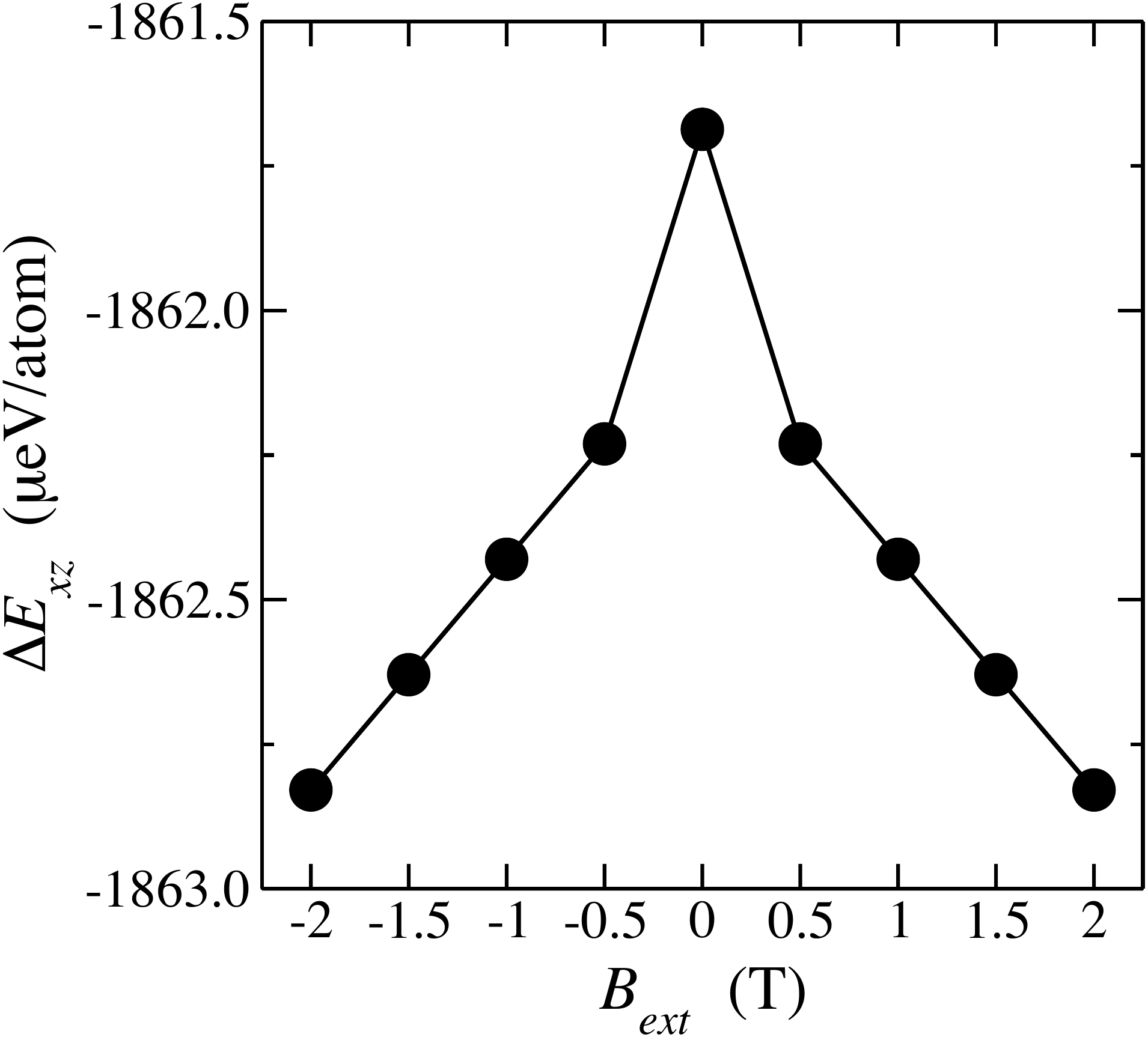}
\end{center}
\caption{Variation of the band energy part of the magnetic anisotropy energy
of a single monolayer of Fe on Ir(111) as function of an applied (constant)
external magnetic field.}%
\label{fig9}%
\end{figure}

Up-to-now the case of using Ir(111) as substrate is much less studied. From a
spin-polarized STM study of (incomplete) monolayers of Fe on Ir(111) it was
found \cite{Bergmann} that the stacking of the Fe atoms seemed to be of
fcc-type. Furthermore, by applying an external magnetic field to the tip (and
of course also to the sample) from the impact of this field Bergmann et al.
\cite{Bergmann} concluded that ``the observed superstructure is of magnetic
origin with an out-of-plane magnetization". In order to check whether or not
an externally applied field can be the cause for a possible misinterpretation
of experimental results, the effect of such a field on the band energy part of
the magnetic anisotropy energy was simulated by considering
(non-selfconsistently) a Kohn-Sham-Dirac Hamiltonian with the total
magnetization pointing (a) along the surface normal,
\[
H(\mathbf{r})=c\mathbf{\alpha\cdot p+}\beta mc^{2}+V^{eff}(\mathbf{r)I}
_{4}+\beta\Sigma_{z}\left[  B_{z}^{eff}(\mathbf{r})+B^{ext}\right]  \ ,
\]
and (b) in-plane. In the above equation $\mathbf{\alpha}$ and $\beta$ are
Dirac matrices, $\mathbf{\Sigma}_{z}$ is the $z$-component of the so-called
spin operator, $\mathbf{I}_{4}$ is a four-dimensional unit matrix, and
$B^{ext}$ is a small constant external (magnetic) field. This simulation is
displayed in Fig.~\ref{fig9} and very clearly shows that the fields applied in
experiment most likely only of marginally affect the size of the magnetic
anisotropy. Possible sources of discrepancies between the weak in-plane
anisotropy found in our calculations and the out-of-plane magnetization
reported in \cite{Bergmann} can be of quite different origin. For a smooth,
two-dimensional translationally invariant Fe overlayer on Ir(111) the
anisotropy energy is only slightly negative. Therefore layer relaxations
as well as the finite size of the sample might be of importance.
Complicated geometrical distortions due to the incompleteness of the atomic
layers are very difficult to take into account theoretically. Yet another
possibility is that the experimentally found out-of-plane magnetization
in Fe/Ir(111) is not caused by a perpendicular anisotropy but rather by 
other factors like, e.g., a complicated chiral rotation of the magnetization 
due to higher order exchange interactions \cite{Vedmedenko2007}.
It might even turn out that perfect monolayers and single
adatoms such as considered in here cannot reflect sufficiently well the actual
situation mapped in a particular experiment. 
\\
\\
\textbf{Acknowledgement:}
One of us (CE) wants to acknowledge financial support from the Austrian
Science Foundation (FWF, Project Number W004). The present investigation was
also supported financially by the Oak Ridge National Laboratory (Subcontract
Nr. 40 000 43271).Financial support by the Deutsche Forschungsgemeinschaft 
via SFB 668 Magnetismus vom Einzelatom zur Nanostruktur is gratefully acknowledged.


\begin{thebibliography}{99}
\bibitem {afm}In the special case of a single mononolayer of Fe on Ir(111) the
ferromagnetic state of the system is energetically favored with respect to the
antiferromagnetic state only by a total enegy difference of $-0.021$ meV, the
differences in the magnetic moments being of the order of $10^{-7}-10^{-8}$
$\mu_{B}$/atom.
\end{thebibliography}
\end{document}